\documentclass[10pt,a4paper,onecolumn]{article}
\usepackage{marginnote}
\usepackage{graphicx}
\usepackage{xcolor}
\usepackage{authblk,etoolbox}
\usepackage{titlesec}
\usepackage{calc}
\usepackage{tikz}
\usepackage{hyperref}
\hypersetup{colorlinks,breaklinks,
            urlcolor=[rgb]{0.0, 0.5, 1.0},
            linkcolor=[rgb]{0.0, 0.5, 1.0}}
\usepackage{caption}
\usepackage{tcolorbox}
\usepackage{amssymb,amsmath}
\usepackage{ifxetex,ifluatex}
\usepackage{seqsplit}
\usepackage{xstring}

\usepackage{float}
\let\origfigure\figure
\let\endorigfigure\endfigure
\renewenvironment{figure}[1][2] {
    \expandafter\origfigure\expandafter[H]
} {
    \endorigfigure
}

\usepackage{fixltx2e} 
\usepackage[
  backend=biber,
]{biblatex}
\bibliography{paper.bib}

\let\textttOrig=\texttt
\def\texttt#1{\expandafter\textttOrig{\seqsplit{#1}}}
\renewcommand{\seqinsert}{\ifmmode
  \allowbreak
  \else\penalty6000\hspace{0pt plus 0.02em}\fi}

\makeatletter
\let\href@Orig=\href
\def\href@Urllike#1#2{\href@Orig{#1}{\begingroup
    \def\Url@String{#2}\Url@FormatString
    \endgroup}}
\def\href@Notdoi#1#2{\def\tempa{#1}\def\tempb{#2}%
  \ifx\tempa\tempb\relax\href@Urllike{#1}{#2}\else
  \href@Orig{#1}{#2}\fi}
\def\href#1#2{%
  \IfBeginWith{#1}{https://doi.org}%
  {\href@Urllike{#1}{#2}}{\href@Notdoi{#1}{#2}}}
\makeatother

\usepackage[top=3.5cm, bottom=3cm, right=1.5cm, left=1.0cm,
            headheight=2.2cm, reversemp, includemp, marginparwidth=4.5cm]{geometry}

\titleformat{\section}
  {\normalfont\sffamily\Large\bfseries}
  {}{0pt}{}
\titleformat{\subsection}
  {\normalfont\sffamily\large\bfseries}
  {}{0pt}{}
\titleformat{\subsubsection}
  {\normalfont\sffamily\bfseries}
  {}{0pt}{}
\titleformat*{\paragraph}
  {\sffamily\normalsize}

\usepackage{fancyhdr}
\makeatletter
\let\ps@plain\ps@fancy
\fancyheadoffset[L]{4.5cm}
\fancyfootoffset[L]{4.5cm}

\definecolor{linky}{rgb}{0.0, 0.5, 1.0}

\newtcolorbox{repobox}
   {colback=red, colframe=red!75!black,
     boxrule=0.5pt, arc=2pt, left=6pt, right=6pt, top=3pt, bottom=3pt}

\newcommand{\ExternalLink}{%
   \tikz[x=1.2ex, y=1.2ex, baseline=-0.05ex]{%
       \begin{scope}[x=1ex, y=1ex]
           \clip (-0.1,-0.1)
               --++ (-0, 1.2)
               --++ (0.6, 0)
               --++ (0, -0.6)
               --++ (0.6, 0)
               --++ (0, -1);
           \path[draw,
               line width = 0.5,
               rounded corners=0.5]
               (0,0) rectangle (1,1);
       \end{scope}
       \path[draw, line width = 0.5] (0.5, 0.5)
           -- (1, 1);
       \path[draw, line width = 0.5] (0.6, 1)
           -- (1, 1) -- (1, 0.6);
       }
   }

\patchcmd{\@maketitle}{center}{flushleft}{}{}
\patchcmd{\@maketitle}{center}{flushleft}{}{}
\patchcmd{\@maketitle}{\LARGE}{\LARGE\sffamily}{}{}
\def\maketitle{{%
  
  \AB@maketitle}}
\makeatletter
\renewcommand\AB@affilsepx{ \protect\Affilfont}
\renewcommand\AB@affilnote[1]{{\bfseries #1}\hspace{3pt}}
\renewcommand{\affil}[2][]%
   {\newaffiltrue\let\AB@blk@and\AB@pand
      \if\relax#1\relax\def\AB@note{\AB@thenote}\else\def\AB@note{#1}%
        \setcounter{Maxaffil}{0}\fi
        \begingroup
        \let\href=\href@Orig
        \let\texttt=\textttOrig
        \let\protect\@unexpandable@protect
        \def\thanks{\protect\thanks}\def\footnote{\protect\footnote}%
        \@temptokena=\expandafter{\AB@authors}%
        {\def\\{\protect\\\protect\Affilfont}\xdef\AB@temp{#2}}%
         \xdef\AB@authors{\the\@temptokena\AB@las\AB@au@str
         \protect\\[\affilsep]\protect\Affilfont\AB@temp}%
         \gdef\AB@las{}\gdef\AB@au@str{}%
        {\def\\{, \ignorespaces}\xdef\AB@temp{#2}}%
        \@temptokena=\expandafter{\AB@affillist}%
        \xdef\AB@affillist{\the\@temptokena \AB@affilsep
          \AB@affilnote{\AB@note}\protect\Affilfont\AB@temp}%
      \endgroup
       \let\AB@affilsep\AB@affilsepx
}
\makeatother

\renewcommand\Affilfont{\sffamily\small\mdseries}
\ifnum 0\ifxetex 1\fi\ifluatex 1\fi=0 
  \usepackage[T1]{fontenc}
  \usepackage[utf8]{inputenc}

\else 
  \ifxetex
    \usepackage{mathspec}
  \else
    \usepackage{fontspec}
  \fi
  \defaultfontfeatures{Ligatures=TeX,Scale=MatchLowercase}

\fi
\IfFileExists{upquote.sty}{\usepackage{upquote}}{}
\IfFileExists{microtype.sty}{%
\usepackage{microtype}
\UseMicrotypeSet[protrusion]{basicmath} 
}{}

\usepackage{hyperref}
\hypersetup{unicode=true,
            pdftitle={Asimov: A framework for coordinating parameter estimation workflows},
            pdfborder={0 0 0},
            breaklinks=true}
\let\addcontentslineOrig=\addcontentsline
\def\addcontentsline#1#2#3{\bgroup
  \let\texttt=\textttOrig\addcontentslineOrig{#1}{#2}{#3}\egroup}
\let\markbothOrig\markboth
\def\markboth#1#2{\bgroup
  \let\texttt=\textttOrig\markbothOrig{#1}{#2}\egroup}
\let\markrightOrig\markright
\def\markright#1{\bgroup
  \let\texttt=\textttOrig\markrightOrig{#1}\egroup}

\usepackage{graphicx,grffile}
\makeatletter
\def\maxwidth{\ifdim\Gin@nat@width>\linewidth\linewidth\else\Gin@nat@width\fi}
\def\maxheight{\ifdim\Gin@nat@height>\textheight\textheight\else\Gin@nat@height\fi}
\makeatother
\setkeys{Gin}{width=\maxwidth,height=\maxheight,keepaspectratio}
\IfFileExists{parskip.sty}{%
\usepackage{parskip}
}{
\setlength{\parindent}{0pt}
\setlength{\parskip}{6pt plus 2pt minus 1pt}
}
\ifx\paragraph\undefined\else
\let\oldparagraph\paragraph
\renewcommand{\paragraph}[1]{\oldparagraph{#1}\mbox{}}
\fi
\ifx\subparagraph\undefined\else
\let\oldsubparagraph\subparagraph
\renewcommand{\subparagraph}[1]{\oldsubparagraph{#1}\mbox{}}
\fi

\title{Asimov: A framework for coordinating parameter estimation workflows}

        \author[1]{Daniel Williams}
          \author[1]{John Veitch}
          \author[4]{Maria Luisa Chiofalo}
          \author[6]{Patricia Schmidt}
          \author[5]{Richard P. Udall}
          \author[2, 3]{Avi Vajpeji}
          \author[7]{Charlie Hoy}
    
      \affil[1]{School of Physics and Astronomy, University of Glasgow, Glasgow, G12
8QQ, United Kingdom}
      \affil[2]{School of Physics and Astronomy, Monash University, Clayton VIC 3800,
Australia}
      \affil[3]{OzGrav: The ARC Centre of Excellence for Gravitational Wave Discovery,
Clayton VIC 3800, Australia}
      \affil[4]{Department of Physics ``Enrico Fermi'', University of Pisa, and INFN,
Largo Bruno Pontecorvo 3 I-56126 Pisa, Italy}
      \affil[5]{LIGO Laboratory, California Institute of Technology}
      \affil[6]{School of Physics and Astronomy and Institute for Gravitational Wave
Astronomy, University of Birmingham, Edgbaston, Birmingham, B15 9TT,
United Kingdom}
      \affil[7]{Cardiff University, Cardiff CF24 3AA, UK}
  \date{\vspace{-5ex}}

\begin{document}
\maketitle

\marginpar{
  \sffamily\small

  {\bfseries DOI:} \href{https://doi.org/}{\color{linky}{}}

  \vspace{2mm}

  {\bfseries Software}
  \begin{itemize}
    \setlength\itemsep{0em}
    \item \href{https://github.com/openjournals/joss-reviews/issues/4170}{\color{linky}{Review}} \ExternalLink
    \item \href{https://git.ligo.org/asimov/asimov}{\color{linky}{Repository}} \ExternalLink
    \item \href{https://doi.org/10.5281/zenodo.4024432}{\color{linky}{Archive}} \ExternalLink
  \end{itemize}

  \vspace{2mm}

  {\bfseries Submitted:} \\
  {\bfseries Published:} 

  \vspace{2mm}
  {\bfseries License}\\
  Authors of papers retain copyright and release the work under a Creative Commons Attribution 4.0 International License (\href{https://creativecommons.org/licenses/by/4.0/}{\color{linky}{CC BY 4.0}}).
}

\hypertarget{summary}{%
\section{Summary}\label{summary}}

Since the first detection in 2015 of gravitational waves from compact
binary coalescence (B. Abbott and others 2016a), improvements to the
Advanced LIGO and Advanced Virgo detectors have expanded our view into
the universe for these signals. Searches of the of the latest observing
run (O3) have increased the number of detected signals to 90, at a rate
of approximately 1 per week (The LIGO Scientific Collaboration, the
Virgo Collaboration, Abbott, et al. 2021; The LIGO Scientific
Collaboration, the Virgo Collaboration, the KAGRA Collaboration, et al.
2021). Future observing runs are expected to increase this even
further(Abbott and others 2020). Bayesian analysis of the signals can
reveal the properties of the coalescing black holes and neutron stars by
comparing predicted waveforms to the observed data (B. Abbott and others
2016b). The proliferating number of detected signals, the increasing
number of methods that have been deployed (Veitch and others 2015;
Ashton and others 2019; Lange, O'Shaughnessy, and Rizzo 2018), and the
variety of waveform models (Ossokine and others 2020; Khan et al. 2020;
Pratten and others 2021) create an ever-expanding number of analyses
that can be considered.

\texttt{Asimov} is a python package which is designed to simplify and
standardise the process of configuring these analyses for a large number
of events. It has already been used in developing analyses in three
major gravitational wave catalog publications (Abbott and others 2021;
The LIGO Scientific Collaboration, the Virgo Collaboration, Abbott, et
al. 2021; The LIGO Scientific Collaboration, the Virgo Collaboration,
the KAGRA Collaboration, et al. 2021). The source code of
\texttt{Asimov} is archived to Zenodo (Williams et al. 2021).

\hypertarget{statement-of-need}{%
\section{Statement of Need}\label{statement-of-need}}

While these developments are positive, they also bring considerable
challenges. The first of these lies with the high rate at which
gravitational waves can now be detected; thanks to the improved
sensitivity of the detectors they observe a much larger volume of space,
and the increasing size of the detector network has also increased the
total time during which observations occur. The second comes from
developments in the analysis techniques and related software.
Development of these techniques has accelerated in a short period of
time, and the landscape of analysis software has become diverse. It is
desirable to be able to use these techniques with ease, but thanks to
the highly distributed development process which has produced them, they
often have highly heterogeneous interfaces.

We developed asimov as a solution to both of these problems, as it is
capable both of organising and tracking a large number of on-going
analyses, but also of performing setup and post-processing of several
different analysis pipelines, providing a single uniform interface. The
software has been designed to be easily extensible, making integration
with new pipelines straight-forward.

In addition, ensuring that the large number of analyses are completed
successfully, and their results collated efficiently proved a formidable
challenge when relying on ``by-hand'' approaches. The LIGO Scientific
Collaboration operate a number of high-throughput computing facilities
(the LIGO Data Grid {[}LDG{]}) which are themselves controlled by the
\texttt{htcondor} scheduling system. \texttt{asimov} monitors the
progress of jobs within the \texttt{htcondor} ecosystem, resubmits jobs
to the cluster which fail due to transient problems, such as file I/O
errors in computing nodes, and detects the completion of analysis jobs.
Upon completion of a job the results are post-processed using the
\texttt{PESummary} python package (Hoy and Raymond 2021), and humans can
be alerted by a message posted by \texttt{asimov} to a Mattermost or
Slack channel. Interaction with \texttt{htcondor} will also allow jobs
to be submitted to the Open Science Grid in the future.

Prior to the development of \texttt{asimov} analyses of gravitational
wave data had been configured and run manually, or had relied on
collections of shell scripts. \texttt{Asimov} therefore constitutes an
new approach, designed to be both more maintainable, and to improve the
reproduciblity of results generated by analysis pipelines.

\hypertarget{implementation}{%
\section{Implementation}\label{implementation}}

In order to produce a uniform interface to all of its supported
pipelines, \texttt{asimov} implements a YAML-formatted configuration
file, which is referred to as its ``production ledger''. This file is
used to specify the details of each event to be analysed, details about
the data sources, and details of each pipeline which should be applied
to the specified data. This allows identical settings to be used with
multiple different pipelines, with a minimum of configuration, reducing
the possibility of transcription errors between setups. In the current
implementation of \texttt{Asimov} the production ledger is stored using
an issue tracker on a custom \texttt{Gitlab} instance, with each issue
representing a different event. This approach is, however, neither
flexible nor scalable, and future development will use an alternative
means of storing the ledger.

\texttt{Asimov} simplifies the process of gathering and collating the
various settings and data-products required to configure an analysis.
These include data quality information: data from gravitational wave
detectors can be affected by non-stationary noise or ``glitches'' which
must be either be removed before analysis, or the analysis must be
configured to mitigate their effect on final results. These data are
provided to \texttt{Asimov} in YAML format from the appropriate team,
and used to make appropriate selections in the analysis.

The analysis of gravitational wave data is generally performed within a
Bayesian framework, which requires prior probability distributions being
chosen before the analysis. Ideally these distributions would be chosen
such that a very broad range of parameter values is explored and
sampled, however this is computationally impractical, and to improve the
speed and efficiency of the analysis a rough guess of the parameters is
required. This is normally determined by ``preliminary'' analyses,
rougher, rapid analyses performed, which are themselves informed by the
detection process which identified the event in the raw detector data.
These prior data are analysed by the \texttt{PEConfigurator} tool to
determine appropriate prior ranges, and settings for the waveform
approximant to be used in the analysis.

The calibration of the detectors; the correspondance between the strain
on the detector and the intensity of light at the interferometer's exit
port, can change over the course of an observing run. The uncertainty in
this quantity is marginalised by many of the analyses, which requires
data files to be collected and provided to the analyses.

Once the correct data, settings, and calibration information has been
identified and collected it is possible to configure analyses.
\texttt{Asimov} allows analyses to be described as a dependency tree,
allowing the output data products from one analysis to be used as an
input for another. This is often useful for coordinating the
determination of the PSD of the analysed data.

Each pipeline is configured with a mixture of configuration files and
command-line arguments. \texttt{Asimov} produces the
appropriately-formatted configuration file for each pipeline using a
template and substitutions from the production ledger. The appropriate
command line program is then run for the given pipeline, in order to
produce an execution environment and submission data for the
\texttt{htcondor} scheduling system. This is then submitted to the LDG,
and the job id is collected and stored by \texttt{asimov}.

It is then possible to automatically monitor the progress of jobs on the
LDG, produce a webpage summarising the status of all on-going analyses,
and detect the completion of jobs and initialise post-processing.

\hypertarget{acknowledgements}{%
\section{Acknowledgements}\label{acknowledgements}}

DW and JV acknowledge support from Science and Technology Facilities
Council (STFC) grants ST/V001736/1 and ST/V005634/1. PS acknowledges
support from STFC grant ST/V005677/1. CH acknowledges support from STFC
grant ST/N005430/1 and European Research Council (ERC) Consolidator
Grant 647839. We are grateful for the support of our colleagues in the
LIGO-Virgo Compact Binary Coalescence Parameter Estimation working
group, including, but not limited to Christopher Berry for his
suggestions and insights. The authors are grateful for computational
resources provided by the Leonard E Parker Center for Gravitation,
Cosmology and Astrophysics at the University of Wisconsin-Milwaukee and
supported by National Science Foundation Grants PHY-1626190 and
PHY-1700765.

\hypertarget{references}{%
\section*{References}\label{references}}
\addcontentsline{toc}{section}{References}

\hypertarget{refs}{}
\leavevmode\hypertarget{ref-GW150914}{}%
Abbott, B.P., and others. 2016a. ``Observation of Gravitational Waves
from a Binary Black Hole Merger.'' \emph{Phys. Rev. Lett.} 116 (6):
061102. \url{https://doi.org/10.1103/PhysRevLett.116.061102}.

\leavevmode\hypertarget{ref-TheLIGOScientific:2016wfe}{}%
---------. 2016b. ``Properties of the Binary Black Hole Merger
GW150914.'' \emph{Phys. Rev. Lett.} 116 (24): 241102.
\url{https://doi.org/10.1103/PhysRevLett.116.241102}.

\leavevmode\hypertarget{ref-Abbott:2020qfu}{}%
---------. 2020. ``Prospects for observing and localizing
gravitational-wave transients with Advanced LIGO, Advanced Virgo and
KAGRA.'' \emph{Living Rev. Rel.} 23 (1): 3.
\url{https://doi.org/10.1007/s41114-020-00026-9}.

\leavevmode\hypertarget{ref-GWTC2}{}%
Abbott, R., and others. 2021. ``GWTC-2: Compact Binary Coalescences
Observed by LIGO and Virgo During the First Half of the Third Observing
Run.'' \emph{Phys. Rev. X} 11: 021053.
\url{https://doi.org/10.1103/PhysRevX.11.021053}.

\leavevmode\hypertarget{ref-Ashton:2018jfp}{}%
Ashton, Gregory, and others. 2019. ``BILBY: A user-friendly Bayesian
inference library for gravitational-wave astronomy.'' \emph{Astrophys.
J. Suppl.} 241 (2): 27. \url{https://doi.org/10.3847/1538-4365/ab06fc}.

\leavevmode\hypertarget{ref-pesummary}{}%
Hoy, Charlie, and Vivien Raymond. 2021. ``PESUMMARY: The code agnostic
Parameter Estimation Summary page builder.'' \emph{SoftwareX} 15 (July):
100765. \url{https://doi.org/10.1016/j.softx.2021.100765}.

\leavevmode\hypertarget{ref-Khan:2019kot}{}%
Khan, Sebastian, Frank Ohme, Katerina Chatziioannou, and Mark Hannam.
2020. ``Including higher order multipoles in gravitational-wave models
for precessing binary black holes.'' \emph{Phys. Rev. D} 101 (2):
024056. \url{https://doi.org/10.1103/PhysRevD.101.024056}.

\leavevmode\hypertarget{ref-Lange:2018pyp}{}%
Lange, Jacob, Richard O'Shaughnessy, and Monica Rizzo. 2018. ``Rapid and
accurate parameter inference for coalescing, precessing compact
binaries.'' \emph{arXiv E-Prints}, May.
\url{http://arxiv.org/abs/1805.10457}.

\leavevmode\hypertarget{ref-Ossokine:2020kjp}{}%
Ossokine, Serguei, and others. 2020. ``Multipolar Effective-One-Body
Waveforms for Precessing Binary Black Holes: Construction and
Validation.'' \emph{Phys. Rev. D} 102 (4): 044055.
\url{https://doi.org/10.1103/PhysRevD.102.044055}.

\leavevmode\hypertarget{ref-Pratten:2020ceb}{}%
Pratten, Geraint, and others. 2021. ``Computationally efficient models
for the dominant and subdominant harmonic modes of precessing binary
black holes.'' \emph{Phys. Rev. D} 103 (10): 104056.
\url{https://doi.org/10.1103/PhysRevD.103.104056}.

\leavevmode\hypertarget{ref-GWTC2.1}{}%
The LIGO Scientific Collaboration, the Virgo Collaboration, R. Abbott,
T. D. Abbott, F. Acernese, K. Ackley, C. Adams, et al. 2021. ``GWTC-2.1:
Deep Extended Catalog of Compact Binary Coalescences Observed by LIGO
and Virgo During the First Half of the Third Observing Run.''
\emph{arXiv E-Prints}, August, arXiv:2108.01045.
\url{http://arxiv.org/abs/2108.01045}.

\leavevmode\hypertarget{ref-GWTC3}{}%
The LIGO Scientific Collaboration, the Virgo Collaboration, the KAGRA
Collaboration, R. Abbott, T. D. Abbott, F. Acernese, K. Ackley, et al.
2021. ``GWTC-3: Compact Binary Coalescences Observed by LIGO and Virgo
During the Second Part of the Third Observing Run.'' \emph{arXiv
E-Prints}, November, arXiv:2111.03606.
\url{http://arxiv.org/abs/2111.03606}.

\leavevmode\hypertarget{ref-lalinference}{}%
Veitch, J., and others. 2015. ``Parameter estimation for compact
binaries with ground-based gravitational-wave observations using the
LALInference software library.'' \emph{Phys. Rev. D} 91 (4): 042003.
\url{https://doi.org/10.1103/PhysRevD.91.042003}.

\leavevmode\hypertarget{ref-zenodo}{}%
Williams, Daniel, Duncan Macleod, Avi Vajpeyi, and James Clark. 2021.
``Transientlunatic/Asimov,'' July.
\url{https://doi.org/10.5281/zenodo.4024432}.

\end{document}